\documentclass[a4paper]{article}

\usepackage{graphicx}

\usepackage{amssymb}

\begin{document}

\begin{center}

\Large{A model for spin-polarized transport in\\perovskite
manganite bi-crystal grain boundaries}

\vspace{0.3cm}

\normalsize{R. Gunnarsson$^1$, A. Kadigrobov$^{1,2}$ and Z.
Ivanov$^1$}

\vspace{0.2cm}

$^1$\textit{Department of Microelectronics and Nanoscience,
Chalmers University of Technology and G\"oteborg University, S-412
96 G\"oteborg, Sweden}

$^2$\textit{B. I. Verkin Institute for Low Temperature Physics \&
Engineering, National Academy of Science of Ukraine, 47 Lenin
Ave., 310164 Kharkov, Ukraine}

\vspace{0.2cm}

\today

\end{center}

\begin{abstract}
We have studied the temperature dependence of low-field
magnetoresistance and current-voltage characteristics of a
low-angle bi-crystal grain boundary junction in perovskite
manganite La$_{2/3}$Sr$_{1/3}$MnO$_3$ thin film. By gradually
trimming the junction we have been able to reveal the non-linear
behavior of the latter. With the use of the relation $M_{GB}
\propto M_{bulk}\sqrt{MR^*}$ we have extracted the grain boundary
magnetization. Further, we demonstrate that the built-in potential
barrier of the grain boundary can be modelled by $V_{bi}\propto
M_{bulk}^2 - M_{GB}^2$. Thus our model connects the
magnetoresistance with the potential barrier at the grain boundary
region. The results indicate that the band-bending at the grain
boundary interface has a magnetic origin.
\end{abstract}

\vspace{0.5cm}

Recent studies have shown that grain boundaries (GBs) in
polycrystalline perovskite manganites, so called
colossal-magnetoresistance (CMR) materials, give raise to a
low-field magnetoresistance.\cite{Hwang1996,Gupta1996} To analyze
the contribution from a single grain boundary on
magnetoresistance, thin film bi-crystal structures have previously
been applied by several research
groups.\cite{Mathur1997,Steenbeck1997,Steenbeck1998,Todd1999,Westerburg1999,Klein1999,Gross2000,Hofener2000,Mathieu2001}

Supported by such studies Evetts \textit{et~al}~\cite{Evetts1998}
have suggested that the low-field magnetoresistance $MR^*$
associated with GBs relates to the magnetization $M_{GB}$ of the
interface region. In their model they assume that the resistivity
$\rho$ depends on the magnetization $M$ as $\rho \propto \rho_0
\exp[-M^2/M_B^2]$ ($M_B$ represents the bulk magnetization) and
hence the magnetoresistance $MR\equiv(\rho_0-\rho)/\rho_0$ can for
the GB be determined as
\begin{eqnarray}
MR^* \propto \frac{M_{GB}^2}{M_B^2} \label{eq:MR_M}
\end{eqnarray}
where $MR^*$ is measured at coercive fields and hence $M_{GB}^2
\ll M_B^2$.

In addition to the low-field magnetoresistance, non-linear
current-voltage (I-V) characteristics has been observed for
manganite GBs.\cite{Todd1999,Westerburg1999,Klein1999,Gross2000}
It has previously been demonstrated for other perovskite oxides
that grain boundaries introduce a band-bending
effect.\cite{Mannhart1997,euroceram1998} Further Furukawa has
demonstrated that for CMR materials the magnetization $M$ induces
a shift in the chemical potential with $\Delta \Phi \propto
M^2$.\cite{Furukawa1997} As a result of the dislocations and
crystalline defects, there is a suppressed magnetic order close to
the GB, and thus there is a built-in potential barrier associated
with the GB region. Hence as discussed by Gross
\textit{et~al}~\cite{Gross2000} this built-in potential barrier
originates from the suppressed magnetic order at the paramagnetic
GB, due to the difference in chemical potential in the grain
boundary region as compared to the bulk.

These two models, by Evetts \textit{et~al}~\cite{Evetts1998} and
by Gross \textit{et~al}~\cite{Gross2000}, describe two different
sides of the influence of the grain boundary, the first one on
spin-polarized transport and the second on the potential barrier
causing the non-linear I-V characteristics. However, until now
there is no model that can explain consistently $MR^*$ and the
band-bending effects observed in perovskite manganites. One
obstacle has been the lack of a full set of data for the
magnetization in the GB region. Here we present a new approach to
the problem. First we argue that the grain boundary magnetization
can be found from the low-field magnetoresistance. We then
successfully have applied a method to model the appearance of the
magnetic potential barrier at the GB. As a result of this study we
can show that the origin of the non-linear current-voltage
characteristics is a built-in potential barrier with a barrier
height that is determined by the difference in square between the
bulk and grain boundary magnetizations.

This study has been performed on a symmetric low-angle bi-crystal
grain boundary. Previous transmission electron studies have shown
that manganite grain boundaries are relatively straight and well
defined.\cite{Gross2000,TEM,Miller2000} As manganite material we
have chosen La$_{2/3}$Sr$_{1/3}$MnO$_3$~(LSMO) which has the
largest one-electron bandwidth as well as the highest $T_C$ among
the ordinary perovskite manganites.\cite{Tokura1999}

\begin{figure}
    \begin{center}
        \includegraphics[height=6.5cm]{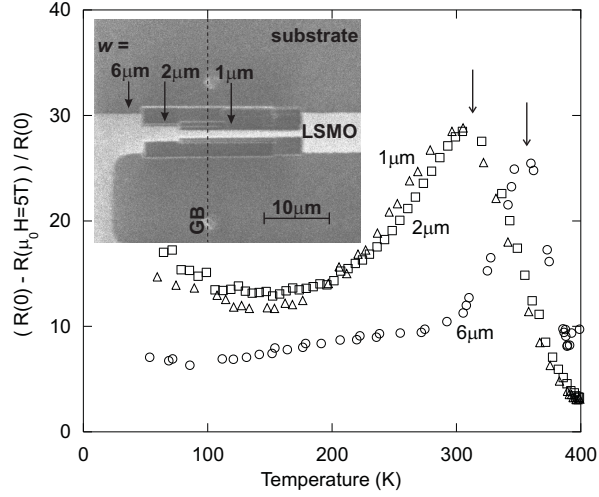}
        \caption{High field (5~T) magnetoresistance as function of temperature.
        The Curie temperatures $T_C$ are indicated with
        arrows. The inset show a secondary-electron image of the GBJ
        geometry after the second trimming process.\label{fig:HFMR}}
    \end{center}
\end{figure}

A LSMO film of the above composition was grown by pulsed laser
deposition on a LaAlO$_3$ bi-crystal substrate with symmetric
misorientation angle of 8.8$^\circ$. In this process a
stoichiometric target was ablated by an excimer laser (KrF,
$\lambda=248$~nm, $\tau=30$~ns) with an energy density of
$\sim1.4$~J/cm$^2$. During deposition the substrate was held at
740$^\circ$C in oxygen pressure of 0.4~mbar. A high degree of
epitaxy of the 90~nm-thick film was verified by X-ray
$\theta-2\theta$ and $\phi-$scans in a four-circle diffractometer.
The film was then patterned with photolithography and Ar-ion
milling into microbridges crossing the GB and forming grain
boundary junctions (GBJs) with a width of $w=6\mu$m. The
magnetoresistance properties were measured in a helium cryostat
with a variable temperature insert ($2$~K$-400$~K) and a 5~T
superconducting magnet. In all measurements the magnetic field was
in the plane of the film and parallel to the GB, the resistance
was measured with a bias current of 10~$\mu$A. The high-field MR
has been deduced from the zero-field and field cooled resistance
measurements, while the low-field MR was measured at stable
temperatures. Current-voltage characteristics at zero external
field were obtained by applying dc current and measuring voltage
in a four point contact geometry. A single GBJ was trimmed, which
revealed the non-linear effect on the I-V curves. The trimming was
performed by a focused ion-beam (FIB) with Ga-ion source. The
junction was trimmed in two steps, the first FIB-process left a
grain boundary junction 2~$\mu$m wide and 20~$\mu$m long, while in
the second FIB-process the junction was trimmed down to
$w=$1~$\mu$m. The junction geometry after the last trimming step
is shown in Fig.~\ref{fig:HFMR}.

The Curie temperature $T_C$ of the sample before trimming was
about 350~K.\cite{howTc} After trimming $T_C$ decreased to about
315~K which could be observed in the resistance and the high-field
MR curves (as indicated by the arrows in Fig.~\ref{fig:HFMR}). The
shift may in part be explained by Ga contamination from the
FIB-process, as other studies have shown that for manganite
materials a 6\% change in Ga-doping may cause a decrease of $T_C$
with 50~K.\cite{Veera2001} An other reason could of course be
structural disorder mainly at the edges of the junction. However
the variation in resistance of the narrower junctions are not more
than would be expected from geometric considerations. Further
trimming of the GBJ (from 2~$\mu$m to 1~$\mu$m) does not change
the transition temperature. Moreover the magnitude of the
high-field (5~T) MR is not affected, even though a broadening in
the peak can be seen. As the resistance and high-field
magnetoresistance primarily are sensitive to the bulk electrode
magnetization the $T_C$ values can be estimated from that data.
Thereby we have found that the Curie temperatures ($T_C$) are
$350$~K, $315$~K and $315$~K for the $w=$6~$\mu$m, 2~$\mu$m and
1~$\mu$m wide GBJs respective.

In the low-field magnetoresistance we first note that the single
GBJ shows a clear multiple-step resistance, similar to what has
been presented in previous studies (see e.g. Fig.~2 in
Ref.~\cite{Todd1999}). Next we observe that the peak low-field
magnetoresistance occur at a field $H^* \approx 30$~mT close to
the coercive field. We associate the multi-step behavior with
domain wall pinning close to the GB. As the junction was trimmed
the number of pinning points decreased and the number of steps
decreased consistently. For single GBJs the MR pattern was not
reproducible, which made us conclude that the domain walls were
pinned at different points for every magnetic field sweep. Due to
this irreproducibility several scans were made to maximize the
$MR^*$.

From the temperature dependence of the $MR^*$, we can extract the
low-field magnetoresistance onset temperature, i.e. the Curie
temperature of the grain boundary region, $T_C^*$. For the
6~$\mu$m wide GBJ we find $T_C^* \approx 320$~K and for the
narrower GBJs $T_C^* \approx 280$~K (2~$\mu$m) and $T_C^* \approx
260$~K (1~$\mu$m). Thus, there is a decrease in $T_C^*$ as the
width of the GBJ is decreased. At the same time the ratio
${T_C^*}/{T_C}$ decreases slightly from about $0.91$ to $0.84$.

\begin{figure}
    \begin{center}
        \includegraphics[height=6.5cm]{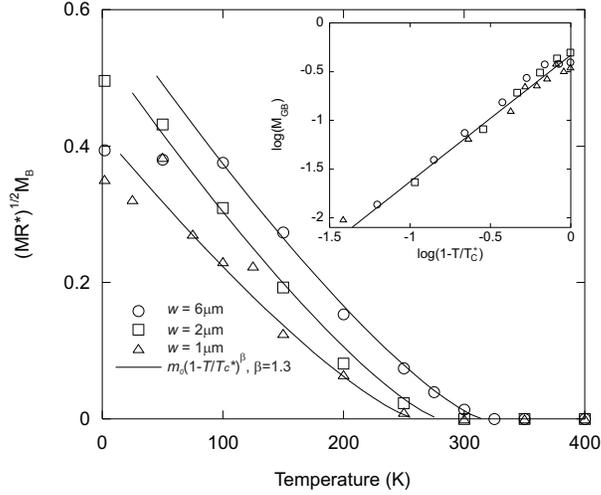}
        \caption{Grain boundary magnetization $M_{GB}$ extracted
        from $M_B \sqrt{MR^*}$ and plotted as function of temperature.
        The lines represent $m_0(1-T/T_C^*)^\beta$ where $\beta=1.3$ is found from the slope of the line
        in the logarithmic plot (the inset). For $w=$~6~$\mu$m, 2~$\mu$m and 1~$\mu$m, $T_C^*=320$~K, 280~K and 260~K and $m_0=$~0.615, 0.54 and 0.42 has been used respectively.
        The data shown correspond to the maximum $MR^*$ values obtained.\label{fig:MRpeak}}
    \end{center}
\end{figure}

The grain boundary magnetization can be found from
Eq.~(\ref{eq:MR_M}) as $M_{GB}= M_B~\sqrt{MR^*}$ except for a
proportionality constant. The bulk electrode magnetization $M_B$
can be obtained from $M_B \propto (1-T/T_C)^\beta$ where
$\beta=0.37$.\cite{Ghosh1998} The general shape of the thereby
extracted $M_{GB}$-curves, shown in Fig.~\ref{fig:MRpeak}, closely
resembles the surface-boundary magnetization measured by Park
\textit{et~al} (Fig. 4 in Ref.~\cite{Park1998b}). From the
estimate of $M_{GB}$ we conclude that the GB region has a certain
amount of spontaneous magnetic order in the whole temperature
range below $T_C^*$, where $T_C^*$ is of the same order as $T_C$.
Moreover in the vicinity of $T_C^*$ we can assume that
magnetization scales as $m \propto (1 - T/T_C^*)^\beta$ where
$\beta$ is the scaling parameter. Such a power-law behavior would
lead to a straight line in a plot of $\log(M_{GB})$ and $\log(1 -
T/T_C^*)$ as in the inset of Fig.~\ref{fig:MRpeak}. Since the
deviation from linear behavior is small in our data we can obtain
(from the slope) a rough estimation of the scaling parameter
$\beta \approx 1.3$.

\begin{figure}
    \begin{center}
        \includegraphics[height=3cm]{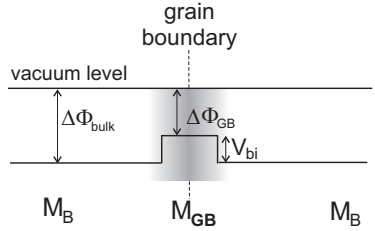}
        \caption{A schematic illustration of the band structure
        close to the GB. The difference in chemical potential shift
        due to difference in magnetization of the bulk and the grain
        boundary region defines the built-in potential $V_{bi}$.\label{fig:Mcartoon}}
    \end{center}
\end{figure}

As $M_{GB}$ is different from $M_S$ we will have different shift
in the chemical potential in bulk electrodes and in the GB region
as illustrated in Fig.~\ref{fig:Mcartoon}. Following the Furukawa
suggestion those shifts will be $\Delta \Phi_{bulk} \propto M_B^2$
and $\Delta \Phi_{GB} \propto M_{GB}^2$. Hence the height of the
built-in potential barrier can be written as
\begin{eqnarray}
    V_{bi} \propto \Delta \Phi_{bulk} - \Delta \Phi_{GB} \propto M_B^2 - M_{GB}^2 = A'~(M_B^2 - M_{GB}^2)\label{eq:Vbi}
\end{eqnarray}
where $A'$ is a proportionality constant.

\begin{figure}
    \begin{center}
        \includegraphics[height=7cm]{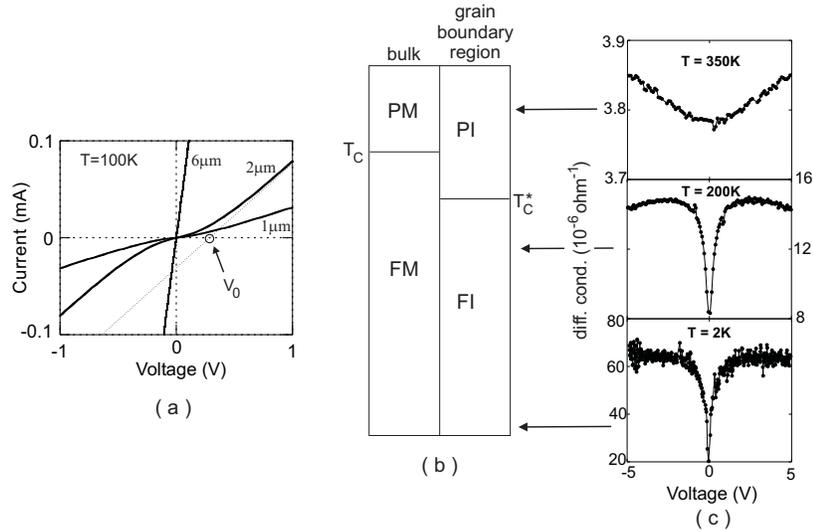}
        \caption{(a) From the linear part of the non-linear I-V curves
        we extrapolate a zero bias-current voltage $V_0$ as
        illustrated here for $T=100$~K. (b) A scetch of the phase
        diagram of the GB region given from the observed $M_{GB}$
        temperature dependence (see Fig.~\ref{fig:MRpeak}) and
        the (c) differential conductance curves. The differential
        conductance is shown for the $w=1$~$\mu$m wide junction at
        $T=350$~K, 200~K and 2~K. PM, FM, PI and FI represent the paramagnetic
        metallic, ferromagnetic metallic, paramagnetic insulating and
        ferromagnetic insulating phases. The transition between PM to FM
        occur at $T_C$ while the PI to FI transition occur at $T_C^*<T_C$.\label{fig:diffcond}}
    \end{center}
\end{figure}

This built-in potential acts a tunnel barrier for the charge
carriers and therefore more information about $V_{bi}$ can be
obtained from the current-voltage characteristics. Several
previous studies (see e.g.
Refs.~\cite{Todd1999,Westerburg1999,Klein1999,Gross2000,Hofener2000})
have shown that the Julliere model~\cite{Julliere1975} for
tunneling transport between ferromagnetic electrodes is not
directly applicable to charge transport across manganite GBs.
Instead charge transfer across this barrier has recently (see
Ref.~\cite{Gross2000}) been described as Glazman-Matveev
transport,\cite{Glazman1988} charge hopping via impurity states in
the interface region, which results in an inelastic contribution
to the conductivity. The current at low voltages in this study
follows $I \propto V^\alpha$, where $\alpha=1 \div 2$, which is
consistent with previously suggested
theories,\cite{Todd1999,Klein1999,Gross2000,Hofener2000} however
due to lack of data points at low-voltages we are not able to
distinguish between the models. At larger voltages, about 1~V, the
I-V curve is fairly linear up to about 5~V where Joule heating
starts to dominate. From this linear part we have extrapolated a
zero bias-current voltage $V_0$ as illustrated in
Fig.~\ref{fig:diffcond}(a). Within the error of a pre-factor we
assume that $V_0$ can give us the temperature dependence of the
built-in potential, $V_{bi}$.\cite{Gunnarsson2001} Thus we have
\begin{eqnarray}
    V_{bi}(T) \propto V_{0}(T) \label{eq:V0}
\end{eqnarray}
by the assumption that the proportionality constant is temperature
independent.

As shown in Fig.~\ref{fig:diffcond}(c) the differential
conductance changes character at $T_C$, however it is still
non-linear for temperatures above. This indicates that the
non-linear behavior of the I-V curves is not solely due to the
difference in magnetization of the GBJ. Above $T_C$ the charge
carriers travel between the metallic paramagnetic
 electrodes via the GB layer which forms a low conductivity metal-insulator-metal (M-I-M) junction. Charge
transport across the GB may still be Glazman-Matveev type even
above $T_C$ since their theory does not account for any dependence
on magnetic state. The difference in the differential conductance
above and below $T_C$ indicates that there is a magnetic
contribution to the band bending. By the shape and the magnitude
of those dependencies it is clear that the major part of the band
bending has a magnetic origin and only an insignificant
contribution comes from the M-I-M junction. Eq.~(\ref{eq:Vbi})
then models the magnetic contribution to the potential barrier of
the GB which is the dominating part below $T_C$.

\begin{figure}
    \begin{center}
        \includegraphics[height=6.5cm]{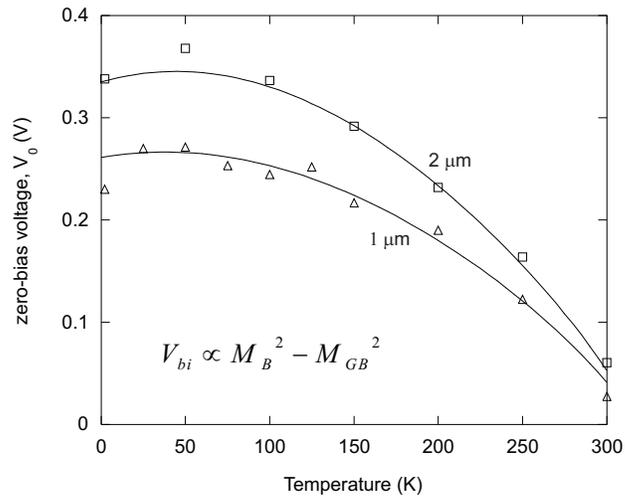}
        \caption{The zero-bias voltage data (open symbols) can
        be mimiced with our model $V_0 = A (M_B^2 - M_{GB}^2)$ (lines). Boxes
        and triangles represent $w=$2~$\mu$m and 1~$\mu$m
        respectively.  Good agreement with experimental data was
        found for $A(w=$2~$\mu$m$)=0.5$ and $A(w=$1~$\mu$m$)=0.39$.\label{fig:V0_T}}
    \end{center}
\end{figure}

Finally merging Eqs. (\ref{eq:Vbi}) and (\ref{eq:V0}) the measured
temperature dependence of the built-in potential can be obtained
from
\begin{eqnarray}
    V_0 = A' (M_B^2 - M_{GB}^2)\label{eq:Vbi_model}
\end{eqnarray}
where $M_B \propto (1-T/T_C)^{0.37}$ and $M_{GB} \propto
(1-T/T_C^*)^{1.3}$. Now $M_{GB}$ is extracted from our measured
data of $MR^*$, thus inserting $C$ as the proportionality between
$MR^*$ and $M_{GB}^2/M_B^2$ in Eq.~(\ref{eq:MR_M}) we will have
\begin{eqnarray}
    V_0 = A ( (1-T/T_C)^{2\cdot0.37} - 1/C~(1-T/T_C^*)^{2\cdot1.3})\label{eq:Vbi_fit}
\end{eqnarray}
where $A$ is related to $A'$ (from Eq.~(\ref{eq:Vbi})). We then
scale this expression (Eq.~(\ref{eq:Vbi_fit})) with the parameter
$A$ to our measured data for $V_0$. The result is shown in
Fig.~\ref{fig:V0_T} for $w=$2~$\mu$m and $w=$1~$\mu$m GBJs
($w=$6~$\mu$m is omitted due to the lack of $V_0$ data). Good
agreement with the data were found for $C=3$ and
$A$(2~$\mu$m$)=0.5$ and $A$(1~$\mu$m$)=0.39$. We note that the
above relation for $M_B$ with $\beta=0.37$ is valid only in the
vicinity of $T_C$ and that the relation for $M_{GB}$ is valid only
in the vicinity of $T_C^*$. However both $M_B$ and $M_{GB}$
describe the general behavior of the magnetization curve for bulk
and the grain boundary region respective in the whole temperature
region. Thus we could use them as a model for magnetization in the
entire temperature range below $T_C$. From Fig.~\ref{fig:V0_T} we
conclude that even though each detail in the $V_0$ data is not
reproduced by our model it does outline the main behavior of the
data.

To summarize we have measured the temperature dependence of the
grain boundary magnetization and the magnetic built-in potential
barrier of the GB when trimming a LSMO low-angle bi-crystal GBJ.
We have thereby employed the proposition of Evetts
\textit{et~al}~\cite{Evetts1998} that $MR^* \propto
M_{GB}^2/M_B^2$. Further we have shown that for low-angle GB the
Curie temperature of the GB region $T_C^*$ is partially suppressed
as compared to the bulk $T_C$ and the ratio $T_C^*/T_C$ is about
0.9. Thus the grain boundary magnetization $M_{GB}$ exists close
to the bulk Curie temperature~\cite{TcCompare}. Moreover we have
been able to extract the behavior of $M_{GB}$ together with a
rough estimate of the scaling parameter $\beta \approx 1.3$.

From the ideas of Gross \textit{et al}~\cite{Gross2000} and
Furukawa~\cite{Furukawa1997} we then have developed a model for
the built-in potential barrier for charge carriers. The model is
based on the different shift in chemical potential due to
difference in magnetization of the GB and the bulk resulting in
$V_{bi}\propto M_B^2 - M_{GB}^2$. We have then shown that the main
features of the temperature dependence of $V_{bi}$ can be
resembled with this model. Moreover we have shown that the
magnetization of the sample has a much higher impact on the
non-linearity of the {I-V} characteristics than the
metal-insulator-metal junction. We attribute this to the high spin
polarization of LSMO.\cite{Park1998a} The observed temperature
dependence of $M_{GB}$ and the differential conductance curves has
allowed us to draw a schematic phase diagram of the GB
(Fig.~\ref{fig:diffcond}(b)) as a complement to ordinary phase
diagrams (see e.g. Ref.~\cite{Tokura1999}). With our model we have
been able to connect the low-field magnetoresistance and the
non-linear behavior observed in GBJs of perovskite manganite
materials.

The work has been supported by The Swedish Research Council (TFR)
and The Board for Strategic Research (SSF) with the programs
"OXIDE" and "Transport in mesoscopic structures".


\begin{thebibliography}{27}
\bibitem{howTc}
The value for $T_C$ is based on superconducting interference
device (SQUID)
  magnetometer measurements on other La$_{2/3}$Sr$_{1/3}$MnO$_3$ on LaAlO$_3$
  bi-crystal samples in the same batch.
\end{thebibliography}

\newpage

\end{document}